\documentclass[conference]{IEEEtran}
\usepackage{cite}
\usepackage[pdftex]{graphicx}
\usepackage[pdftex]{color}
\usepackage[ruled,vlined,linesnumbered]{algorithm2e}
\usepackage[cmex10]{amsmath}
\usepackage{algpseudocode}
\usepackage{array}
\usepackage{url}
\usepackage{mathtools}
\usepackage{amsmath, amssymb}
\usepackage{amsthm}
\usepackage{booktabs}

\theoremstyle{remark}

\theoremstyle{definition}

\DeclarePairedDelimiter\ceil{\lceil}{\rceil}

\begin{document}
\title{Efficient LTE Access with Collision Resolution \\ for Massive M2M Communications}

\author{\IEEEauthorblockN{German Corrales Madue\~no, \v Cedomir Stefanovi\' c, Petar Popovski}
\IEEEauthorblockA{\\Department of Electronic Systems, Aalborg University, Denmark \\
Email: \{gco,cs,petarp\}@es.aau.dk}}
\maketitle

\begin{abstract}
	LTE random access procedure performs satisfactorily in case of asynchronous, uncorrelated traffic arrivals.
	However, when the arrivals are correlated and arrive synchronously, the performance of the random access channel (RACH) is drastically reduced, causing a large number of devices to experience outage.
	In this work we propose a LTE RACH scheme tailored for delay-sensitive M2M services with synchronous traffic arrivals. The key idea is, upon detection of a RACH overload, to apply a collision resolution algorithm based on splitting trees. 
	The solution is implemented on top of the existing LTE RACH mechanism, requiring only minor modifications of the protocol operation and not incurring any changes to the physical layer.
The results are very promising, outperforming the related solutions by a wide margin.
As an illustration, the proposed scheme can resolve $30$k devices with an average of $5$ preamble transmissions and delay of $1.2$ seconds, under a realistic probability of transmissions error both in the downlink and in the uplink.
	
\end{abstract}

\section{Introduction}
\label{sec:introduction}
	Machine-to-Machine (M2M) services span a wide range, including services like car-to-car, smart grid, smart metering, control/monitoring of homes and industry, e-health, traffic control, surveillance, etc.
	Opposed to the typical human-oriented services, M2M services are not driven by data rates, but by the features of availability and reliability.
	However, attaining required availability and reliability of M2M services is not a trivial issue, due to a potentially massive number of devices involved.
	An astonishing 300k devices per cell are foreseen in future M2M scenarios \cite{METIS}, with potentially thousands of them simultaneously trying to access the network.
	Consider the example of smart grid monitoring - in case of a power outage, thousands of smart meters will try to report the failure.
	These messages should be delivered before the battery dies (i.e., last-gasp reporting), setting the reporting deadline to 500~ms \cite{lastGasp}. 
	In such cases, the LTE random access channel (RACH) becomes overloaded by thousands of simultaneous access attempts \cite{surveyRACHLTE}.

	Recently, there has been a large amount of work devoted to investigation of the approaches how to avoid overloading the RACH to protect both network and users against such events.
	One of the initial approaches is to split the preambles used in the RACH for human and M2M communications \cite{splitPreambles}.
	This way human services are not affected, but the major drawback is that the overload problem for M2M services is aggravated, as the number of available preambles is reduced.
	Another approach is to control the RACH load via backoff adjustments, spreading the preamble retransmissions over time and thus attempting to limit the number of collisions.
	However, due to the different nature of human and M2M communications, a valid backoff for former might not be suitable for the latter.
	In \cite{R2113013} specific M2M backoff and class barring parameters are discussed for delay tolerant devices, where the load in the RACH channel is decreased by a factor of 20.
	However, the delay can raise up to 100~s.
	
	On another hand, only a few solutions for delay-sensitive M2M services have been presented so far.
	One of these is the dynamic allocation, where additional RACH resources are allocated when an overload is detected \cite{R2104662}.
	The drawback of this approach is the notification delay of the additional resources availability.
	In LTE, the number of random access opportunities (RAOs) per frame is broadcasted in the system information block 2 (SIB2); it can take up to 512 radio frames, i.e., 5.12~s, before this broadcast is sent \cite{36331}.
	In \cite{Huaweii} a coordinated random access scheme is proposed, where only one or few representatives of every group report the critical information. 
	This is based on the observation that during the congestion period the correlation of messages across devices within a group is very high.
	The drawbacks in this case are the required coordination among devices within the group and the compromised reliability of relying on a few devices per group to successfully report the delay sensitive information.

	In this work we propose a novel approach to deal with massive synchronous access attempts, tailored for delay-sensitive M2M services.
	Contrary to the mainstream solutions that try to avoid collisions by modifying the parameters of the LTE RACH access procedure, we propose use of a \emph{collision resolution} algorithm to resolve synchronous RACH attempts.
	The motivation lies in the observation that when RACH is overloaded by synchronous access attempts, the massive number collisions inevitably occurs and it is more efficient to resolve these collisions instead to waste time and LTE resources by trying to avoid them.
	The basis of the proposed solution is a q-ary tree splitting technique \cite{janssen}, implemented on the top of the existing LTE RACH procedure and activated when RACH overload is detected.
	Apart from the novel idea of using collision resolution in LTE RACH, the paper contributions are also in presentation of the implementation details and demonstration of the efficiency of the proposed approach to achieve a reliable and timely massive synchronous access.

	The rest of the paper is organized as follows.
	Section~\ref{ssub:lte_overview} presents a brief overview of the standard LTE random access.
	Section~\ref{sec:improvements} describes the proposed solution in details.
	Section~\ref{sec:results} demonstrates the performance results.
	Section~\ref{sec:conclusion} concludes the paper.

\section{LTE RACH Overview}
	\label{ssub:lte_overview}
	
		\begin{figure}
		  \centering
		    \includegraphics[width=0.89\columnwidth]{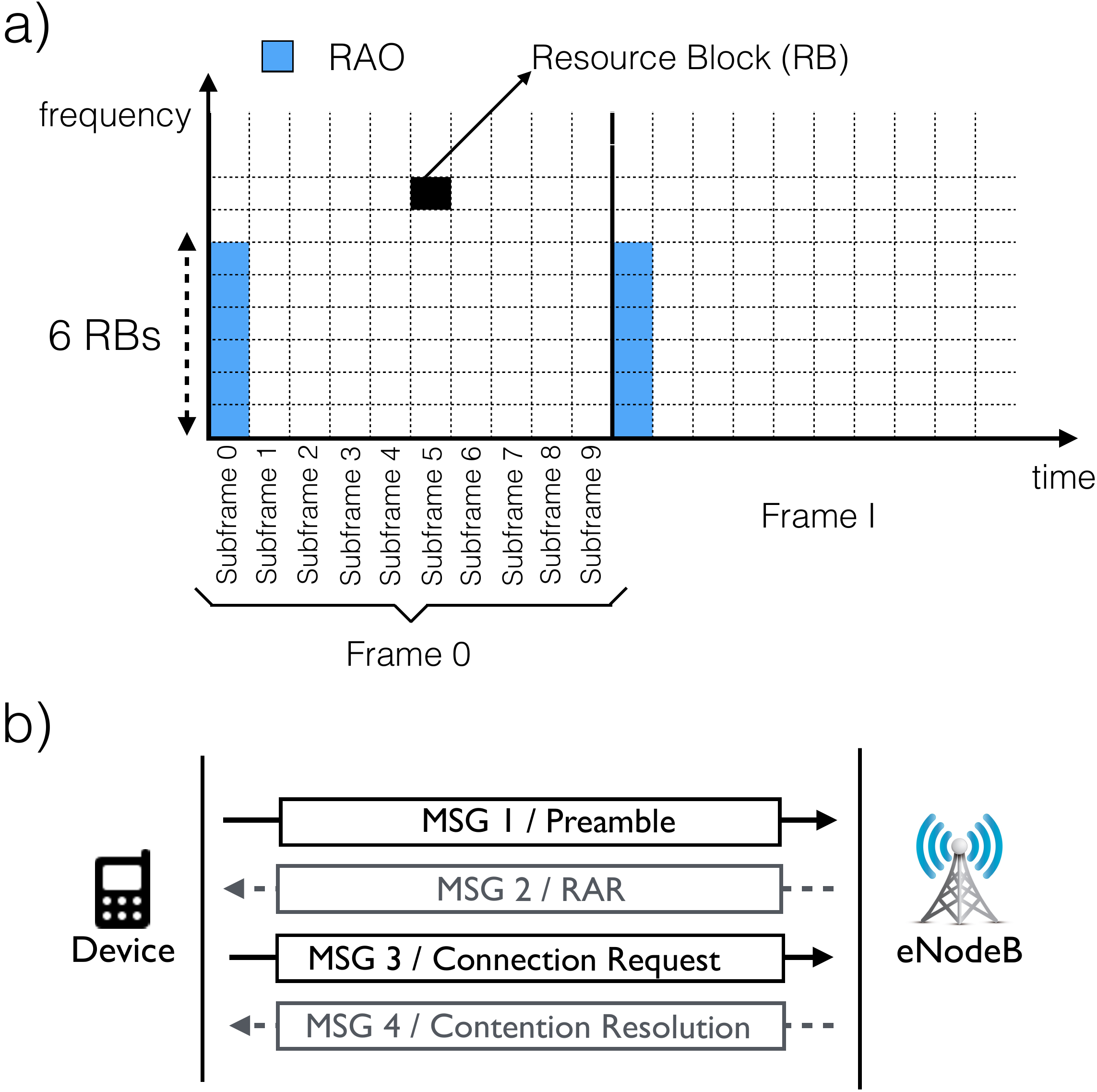}\caption{a) LTE uplink resources with one RAO per frame. b) Message exchange between a device and the eNodeB during the LTE random access procedure.}\label{LTE_grid}
		\end{figure}	

		The uplink resources in LTE for frequency division duplexing (FDD) can be expressed using a 2D grid, see Fig.~\ref{LTE_grid}a), where the x-axis represents time and the y-axis resource blocks (RBs).
		Time is divided in frames, where every frame is composed of ten subframes, and each subframe is of duration $t_s = 1$~ms.
		The amount of RBs per subframe is determined by the available bandwidth in the system, which ranges between 6~RBs and 100~RBs.
		The number of subframes between two consecutive RAOs varies between 1 and 20, where 5 is the most typical value \cite{typical}, providing one RAO every 5~ms.
		Finally, every RAO is composed of 6~RBs, as depicted in Fig.~\ref{LTE_grid}a), and a maximum of one RAO per subframe is allowed.

		The standard LTE random access procedure is of access reservation type, where the devices are contending to reserve resources for their uplink data transmissions using a slotted ALOHA based mechanism. 
		The access procedure comprises exchange of four different messages between a device and the eNodeB, see Fig.~\ref{LTE_grid}b).	
		The first message (MSG~1) consists of a randomly selected preamble sent in the next available RAO.
		There are 64 orthogonal preambles in LTE; some of them are reserved for special purposes and the actual number of available preambles for contention is lower and typically set to 54.
		A typical premise is that the eNodeB can only detect if a preamble has been activated or not, \emph{but not how many devices have actually activated it} \cite{preambleCollision}.
		In other words, if two or more devices send the same preamble in the same RAO, this collision remains undetected. 
		In the next step, the eNodeB replies with the random access response RAR, denoted as MSG~2, to all detected preambles.
		The contending devices monitor the downlink channel, expecting MSG~2 within the next $t_{\mathrm{RAR}}$ seconds.
		If no MSG~2 is received and the maximum of $M$ MSG~1 transmissions is not reached, the random access procedure restarts after a randomly selected time within the interval $t_r\in[0,B]$, where $B$ is a backoff parameter.
		If MSG~2 is received, it includes uplink grant information, pointing to the RB where the connection request (MSG~3) should be sent.
		The connection request indicates the desired operation by the device, such as call/data transmission/measurement report, etc.
		In case when two or more devices activated the same preamble and received the same MSG~2, their MSGs~3 collide in the RB.
		In contrast to the collisions of MSGs~1, collisions of MSGs~3 are detected by the eNodeB.
		The eNodeB replies only to MSGs~3 that did not experience collision, by sending message MSG~4, which allocates the required RBs or denies the request if no resources are available.
		If no MSG~4 is received after $t_{\mathrm{CRT}}$ seconds since MSG~1, the random access procedure is restarted.
		Finally, if after $M$ MSG~1 transmissions a device does not successfully finish all the steps of the random access procedure, an outage is declared. 

		The random access in LTE is well suited for asynchronous arrivals, as a typical RACH configuration offers one RAO with 54 available preambles every 5~ms \cite{typical}, i.e., there are 10.8~k available preambles per second.
		However, as shown in Section~\ref{sec:results}, in case of synchronous traffic arrivals, e.g., alarm events with thousands of devices activated simultaneously, the system cannot cope with the excessive collisions of MSGs~3, and the RACH collapses.
		
\section{The Proposed Solution}
\label{sec:improvements}

	We start by a high level description of the proposed solution.	
	Assume that an event takes place that causes synchronous RACH access attempts by a massive number of devices.
	As the number of contention preambles is limited, the ultimate result is a high number of collided MSGs~3 observed by the eNodeB.\footnote{Note that the eNodeB has only to detect if there is a collision, which could be done in a simple manner, e.g., using an energy detector.}
	This could serve as a trigger for eNodeB to modify the LTE RACH operation, by switching from the slotted ALOHA-based collision avoidance to a collision resolution mechanism.
	Specifically, we propose to use a q-ary tree splitting algorithm \cite{janssen}, leveraging on the LTE orthogonal preambles.
	The notification to the contending devices about the change of RACH operation, as well as direction of the contending devices through the tree splitting, is performed through the feedback messages sent by eNodeB.
	These messages could be implemented by modifying the existing eNodeB messages, as outlined further.
	We proceed by presentation of the details.
	
	
\subsection{LTE RACH Modifications}
	
	\begin{figure*}[t]
	  \centering
	    \includegraphics[width=0.9\textwidth]{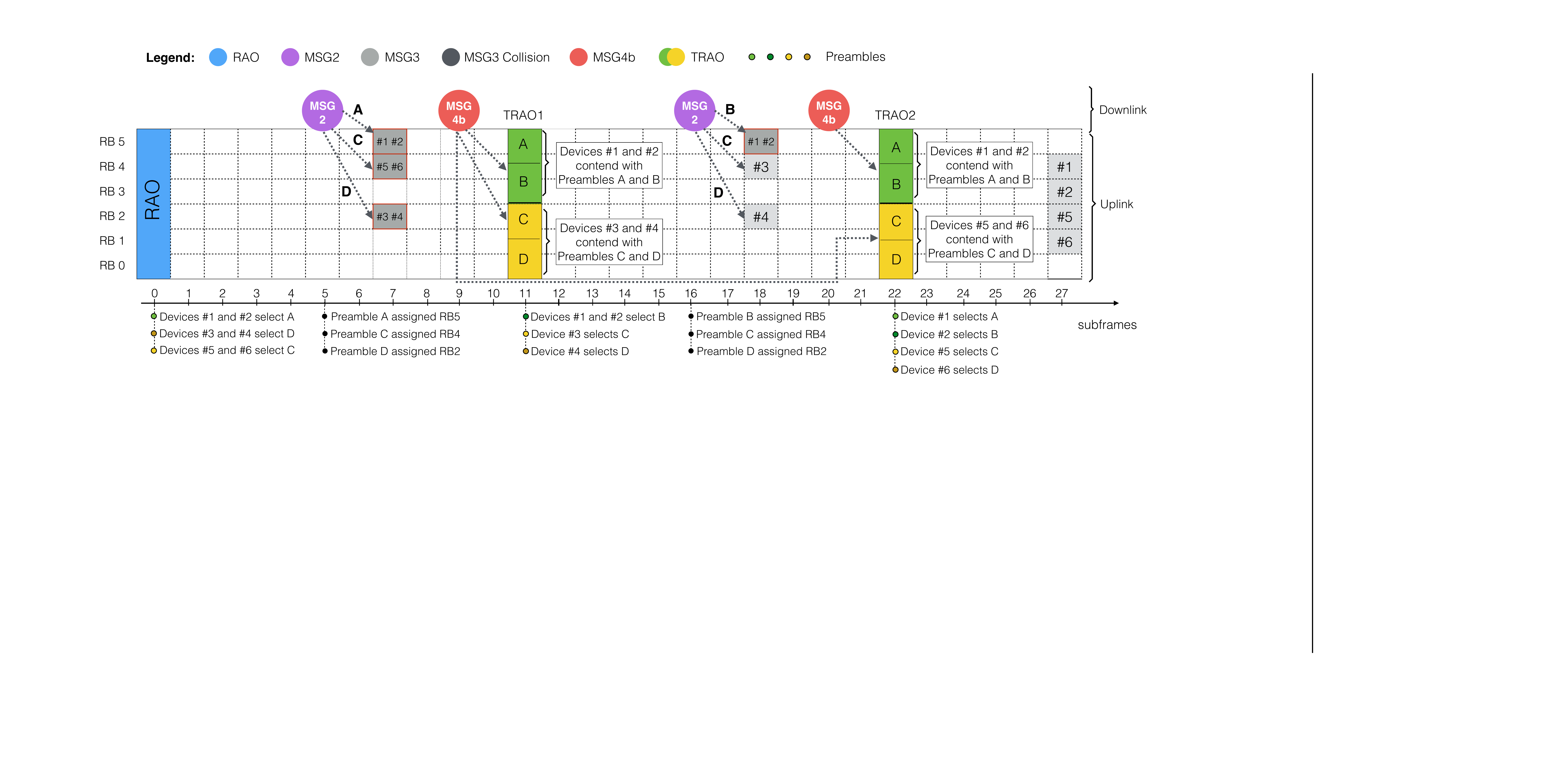}\caption{Illustration of the proposed tree-splitting algorithm.}\label{splitOptions}
	\end{figure*}

	Tree splitting algorithms rely on the use of feedback after every contention attempt; to this end, we propose to use a new type of MSG~4, denoted as MSG~4b.
	Contrary to the standard MSG~4, this message is sent to the devices whose MSGs~3 \emph{collided}, notifying them about the collision and specifying the details of the next contention attempt. 
	Specifically, MSG~4b indicates a set of $q$ preambles to be used for the next contention attempt and the RAO where this contention should take place, denoted as tree-splitting RAO (TRAO).\footnote{We assume that TRAOs are allocated in subframes that are orthogonal to the subframes containing RAOs; thus, the access performance of other services (e.g., human-oriented services) remains unaffected.}
	The recipients of MSG~4b send new MSGs~1, by transmitting a random preamble from the set of $q$ preambles in the designated TRAO, as directed by the eNodeB.
	The eNodeB replies with standard MSG~2 to all detected preambles, and the recipients of MSG~2 send standard MSG~3.
	The eNodeB replies with standard MSG~4 to the non-collided MSGs~3 (i.e., these messages are resolved), and with a new MSG~4b to collided MSGs~3, whose senders continue to participate in the tree-splitting.
	The above procedure repeats until all MSGs~3 are either resolved or the maximum number of preamble transmissions $M$ is reached, when the affected devices declare outage.
	
	For a better understanding we provide an example in Fig.~\ref{splitOptions}, where there are 6 devices and 4 available preambles, denoted as $A$, $B$, $C$, and $D$.
	In subframe 0, devices \#1 and \#2 send preamble $A$, devices \#3 and \#4 send preamble $D$ and devices \#5 and \#6 send preamble $C$.
	The eNodeB detects these three preambles and responds with MSG~2, indicating that MSGs~3 should be sent in subframe 7. When MSGs~3 are transmitted in subframe 7, the collisions are detected and the eNodeB replies with MSGs~4b, indicating that: (i) the devices that sent preamble $A$ should now contend in TRAO in subframe 11 (TRAO1) using preambles $A$ and $B$, (ii) the devices that sent $D$ should also contend in TRAO1 using preambles $C$ and $D$, and (iii) the devices that sent $C$ should contend in TRAO in subframe 22 (TRAO2) with preambles $C$ and $D$.
	Devices \#1 and \#2 again choose the same preamble, their MSGs~3 collide in subframe 18, and they are directed to contend again in TRAO2, using preambles $A$ and $B$.
	This time \#1 and \#2 choose different preambles in TRAO2, so MSGs~3 are allocated different RBs in subframe 27 and do not collide again.
  	Devices \#3 and \#4 choose different preambles already in TRAO1, so their MSGs~3 are resolved in subframe 18.
	Finally, devices \#5 and \#6 choose different preambles in TRAO2, and their MSGs~3 are resolved in subframe~27.
	We also note that for the sake of clarity MSGs~4 are not shown in Fig.~\ref{splitOptions}.
	
	\begin{figure}[t]
	  \centering
	    \includegraphics[width=\columnwidth]{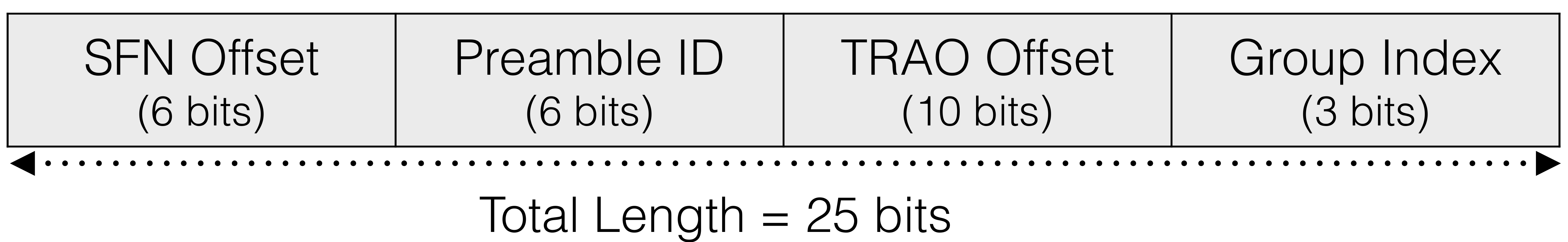}\caption{Proposed MSG~4b format.}\label{fig:formatMSG4b}
	\end{figure}

	A possible format for MSG~4b is depicted in Fig.~\ref{fig:formatMSG4b}.
	The first two fields are used to indicate the devices affected by the message; specifically they indicate the offset in subframe numbers (SFN) between the current SNF and the subframe in which the devices with preamble ID transmitted their MSG~3.
	The last two fields are used for the collision resolution, where TRAO offset and Group Index are used to indicate the SFN in which the TRAO takes place and the group of preambles to be used.
	
	Further, we note that the performance of the random access procedure is also affected by the capacity of the control channel (PDCCH) through which the messages MSG~2, MSG~4 and MSG~4b are sent.
	A straightforward solution is to increase the bandwidth of the system, which indirectly increases the capacity of the PDCCH.
	In this work we consider another approach, proposed in \cite{emergingTech}, where one of the reserved radio network temporal identifiers (RNTI) is dedicated for M2M and defined as M2M-RNTI.
	M2M-RNTI is used by every device to determine who is the recipient of the data or control information.
	If there are not enough resources in the PDCCH, MSG~2, MSGs~4 and MSG~4b for several devices are bundled into one packet data unit and masked with the M2M-RNTI.
	This information is transmitted in the packet data shared-channel (PDSCH), allowing to virtually increase the capacity of the PDCCH.
	Therefore, we assume unlimited downlink capacity, but take into account the amount of required resources when assessing the performance of the proposed solution in Section~\ref{sec:results}.
		
\subsection{Analysis}
\label{sec:analysis}

	\begin{figure}
	  \centering
	    \includegraphics[width=0.75\columnwidth]{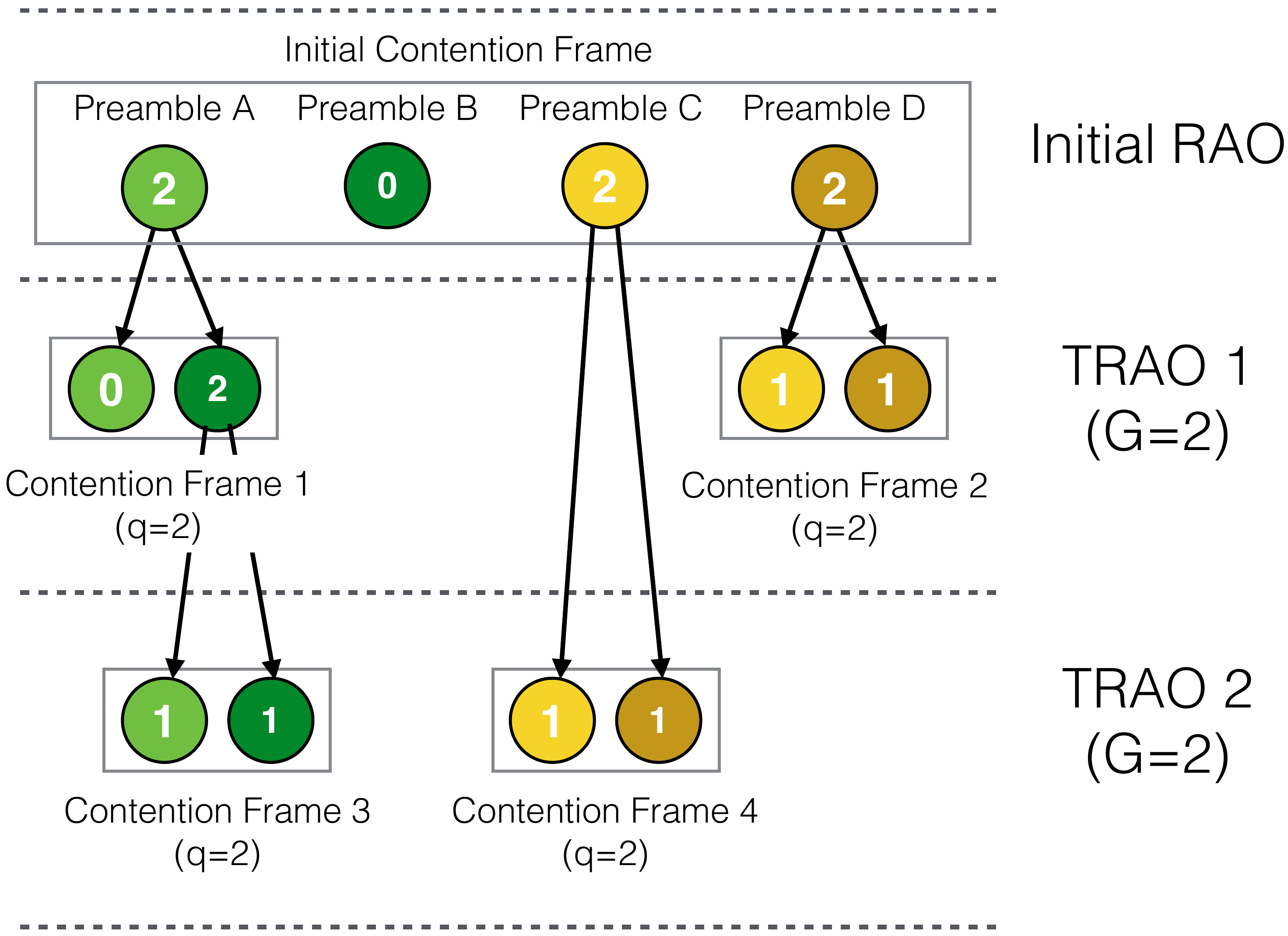}\caption{Illustration of contention resolution with four devices and four preambles.}\label{splitting_tree}
	\end{figure}
	
	In this section we determine the number of transmissions per device, the number of TRAOs required, and the probability of a device exceeding the maximum number of preamble transmissions (outage probability) for the proposed scheme.
	The presented analysis is the adaptation of the one from \cite{janssen}.
	
	The basic structure of the q-ary tree-splitting algorithm is a contention frame, which is composed of $q$ slots.
	Users contend by transmitting in a randomly selected slot; if two or more devices transmit in the same slot, a collision occurs and the slot expands into a new contention frame, again with $q$ slots.
	Every expansion corresponds to a level of the contention tree.
	This procedure repeats until all slots with collisions are resolved.
	
	We translate the above algorithm into LTE RACH terms in the following way.
	The root of the tree is the initial RAO where the original collisions happen, and it constitutes a single contention frame.
	This frame is a exception from all the other frames, as it consists of $N_P$ slots, where the $N_P$ is the total number of available preambles.
	Also, we assume that the set of available preambles is divided in $G$ non-overlapping sets with $q$ preambles in each, i.e., the total number of available preambles is $N_P = G \cdot q$.
	The slots of the initial contention frame that contain collisions are expanded in new contention frames containing $q$ slots each.
	These contention frames take place in TRAOs following the initial RAO; as the available preambles are divided into $G$ sets of $q$ preambles, every TRAO is logically partitioned into $G$ contention frames with $q$ slots in each frame.
	Starting from the slots of the initial contention frame, every subsequent expansion corresponds to a level of the splitting tree; thus, if every slot splits into $q$ new slots, the number of slots in level $m$ is $G q^m$.
		
	Fig.~\ref{splitting_tree} depicts the same example as in Fig.~\ref{splitOptions}, but in the standard tree-splitting representation.
	There are $N_P = 4$ slots in the root contention frame, and $q=2$ slots in all other contention frames; numbers in slots denote how many devices contended in them.
	Note that the contention frames 1, 2 and 4 correspond to the level 2, although they are in different TRAOs, whereas the contention frame 3 corresponds to the level 3, although it is in the same TRAO as contention frame 4. This is due to the fact that every TRAO contains just $G = N_P / q = 2$ contention frames.

	To determine the number of levels, which is equal to the number of preamble transmissions required until MSG~3 is received at eNodeB without collision, we recall the approach from \cite{janssen}.
	We assume that devices in level $m$ are independently and identically randomly distributed over $G q^m$ slots.
	Thus, the probability of only one device transmitting in a slot of level $m$, when there are total of $N \geq 2$ devices at the start of the tree splitting procedure, is:
	\begin{equation}
		P_S(m) = \left(1-\frac{1}{Gq^m}\right)^{N-1}.
	\end{equation}
	The probability that $m$ levels are required to resolve the transmission of the device, denoted by $P_L(m)$, is equal to the probability that the transmission is resolved in level $m$ and it was not resolved in level $m-1$:
	\begin{equation}
		  P_L(m) = P_S(m) - P_S(m-1).
	\end{equation}
	The outage probability of a device, i.e., the probability that more than maximum of $M$ transmissions are required, and the average number of transmissions $T$ are given by:
	\begin{align}
		P_O &= 1 - \sum_{m=1}^M P_L ( m ), \\
		T &= \sum_{k=1}^{\infty} i \cdot P_L ( m ).
	\end{align}
	An approximation to the number of transmissions $T$ can be derived as \cite{janssen}:
	\begin{equation}
		\hat T = \log_m \left( \frac{N-1}{G} \right) - \left( \frac{1}{2} + \frac{\gamma}{\log_m}\right) + \frac{1}{2N \log_m},
		\label{eq:T}
	\end{equation}
	where $\gamma  \approx 0.5772 $ is Euler's constant.
	Further, the number of slots with collisions in level $m$ and therefore the number of contention frames in the next level is given by:
	\begin{equation}
\label{eq:C}
		C(m) = G q^m \left(1- \left(1 -\frac{1}{G q^m}\right)^N\right) - N \left(1 -\frac{1}{G q^m}\right)^{N-1}.
	\end{equation}
	Finally, the expected number of TRAOs required to resolve N devices, denoted as $R$, can be determined from the number of contention frames as:
	\begin{equation}
		R = 1 + \sum_{m=1}^{\infty} \ceil*{\frac{C(m)}{G}},
		\label{eq:R}
	\end{equation}
	where $\lceil \cdot \rceil$ denotes the ceiling function.

\section{Results}
\label{sec:results}

	\begin{figure}
	  \centering
	    \includegraphics[width=0.85\columnwidth]{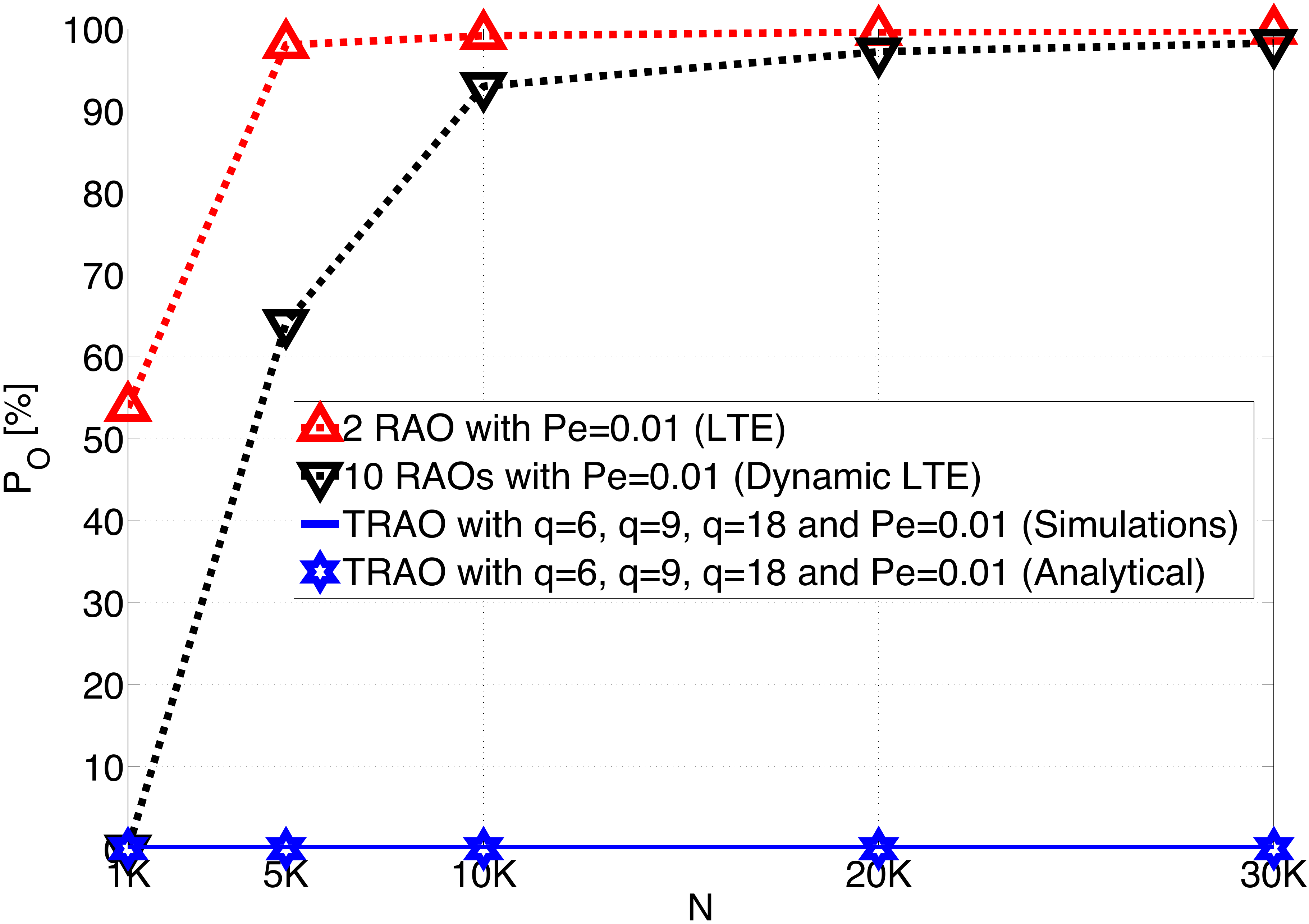}\caption{Outage performance of standard LTE RACH, dynamic allocation and the proposed splitting-tree.}\label{fig:outage}
	\end{figure}
	
	\begin{table}
		\centering
		\begin{tabular}{@{}ll|ll@{}}
			\toprule
			\textbf{Parameter}        	    & \textbf{Value} & \textbf{Parameter}      & \textbf{Value}  \\ \midrule
			Total Number of Preambles ($N_P$)            	& 54  	& MSG~2							& 56 bits   \\ 
			MSG~2 Window ($t_{RAR}$)     	& 5~ms  & MSG~4 							& 20 bits \\ 
			MSG~4 Timer                 	& 24~ms & MSG~4b 							& 25 bits \\
			Maximum Transmissions ($M$) 	& 10  & System BW            	& 20~MHz   \\ 
			Contention Timer ($t_{\mathrm{CRT}}$)	& 48~ms & Backoff ($B_i$)	 & 20~ms \\ 
			eNodeB and UE Processing Time         	& 3~ms &   Modulation 					& QPSK   \\ 
			\bottomrule
		\end{tabular}
	\caption{System Parameters.}
	\label{table:parameters}
	\end{table}

	In this section we present the performance of the proposed access mechanism, obtained both through the analytical approach and simulations.
	We also make a comparison with standard LTE RACH procedure \cite{3GPPTS36.321} and dynamic allocation scheme \cite{R2104662}, whose performances are obtained by simulations.
	For the standard LTE RACH procedure, we use a typical configuration of 2~RAOs per frame\cite{typical}.
	For the dynamic allocation scheme, we assume the maximum of 10~RAOs per frame and that there is no delay to activate the additional RAOs, i.e., we compare our method with the best case of dynamic allocation.
	All the simulations are performed in an event-driven MATLAB simulator, which models the LTE RACH procedure with a probability of error both downlink and uplink of $p_e=0.01$, which is a typical target error rate in LTE control channel \cite{ahmadi2013lte,4394263}.
	The number of simulation repeats is set to 100 for every combination of parameters.
	Since our aim is to compare the performance of the different RACH procedures, we assume that the critical information fits in MSG~3 and no further actions are required; i.e., a device is resolved if MSG~3 is received with no collisions or errors.
	The rest of the parameters of the random access procedure are listed in Table~\ref{table:parameters}; we use a system bandwidth of 20~MHz and note that the similar improvements are observed when less bandwidth is used.
	
	Fig.~\ref{fig:outage} shows the outage probability $P_O$, defined as the percentage of devices not completing the RACH procedure before the maximum number of preamble transmissions $M$ is reached, as function of the number of devices $N$ that synchronously start the random access procedure (i.e., in the same subframe).
	Obviously, a system with 2 RAOs per frame cannot cope with the massive synchronous arrivals and a large percentage of the devices are in outage.
	The dynamic allocation performs better; nevertheless, its performance is worse by a large margin in comparison to the performance of the proposed scheme.
	Specifically, the proposed scheme is able to resolve 30K synchronous attempts for any choice of number of preambles $q$ per contention frame within TRAO with insignificant $P_O$.
	We also note the negligible differences among the results obtained by the analysis and simulations, where the latter include a realistic error probability.
	The same holds for the rest of the presented results.

	\begin{figure}
	  \centering
	    \includegraphics[width=0.83\columnwidth]{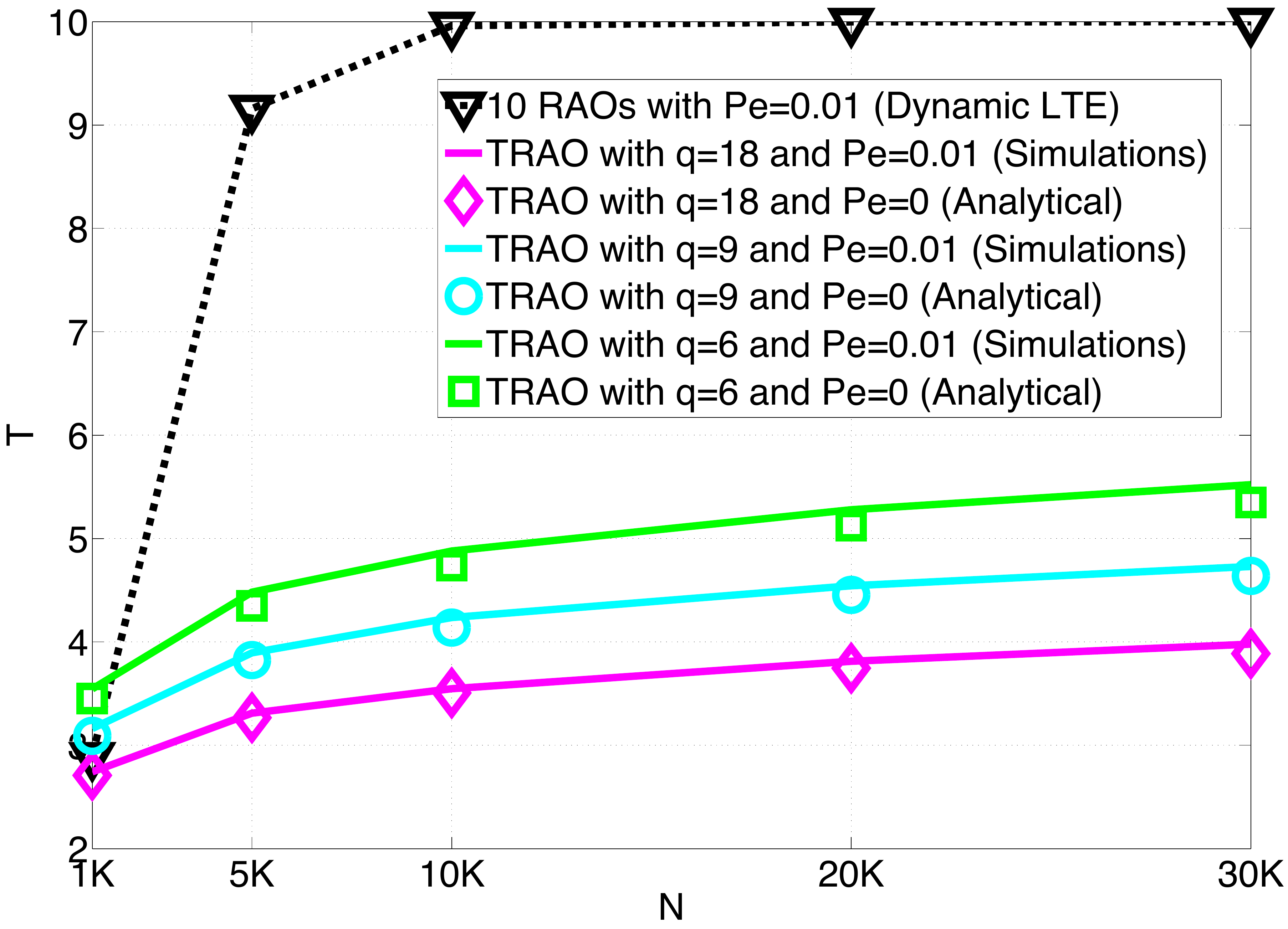}\caption{Average preamble transmissions per device required.}\label{transmissions}
	\end{figure}

	Fig.~\ref{transmissions} shows the average number of preamble transmissions per device $T$ as function of $N$.
	It is clear that 10 preamble transmissions (the allowed maximum $M$) is reached soon by the dynamic procedure, while the proposed scheme requires significantly less preamble transmissions per device.
	Also, the results show that when more preambles $q$ are available to resolve a collision, less preamble transmissions are required on average.
	This could be expected from \eqref{eq:T}, when $ G = N_P / q$ is substituted.

	\begin{figure}[t]
	  \centering
	    \includegraphics[width=0.85\columnwidth]{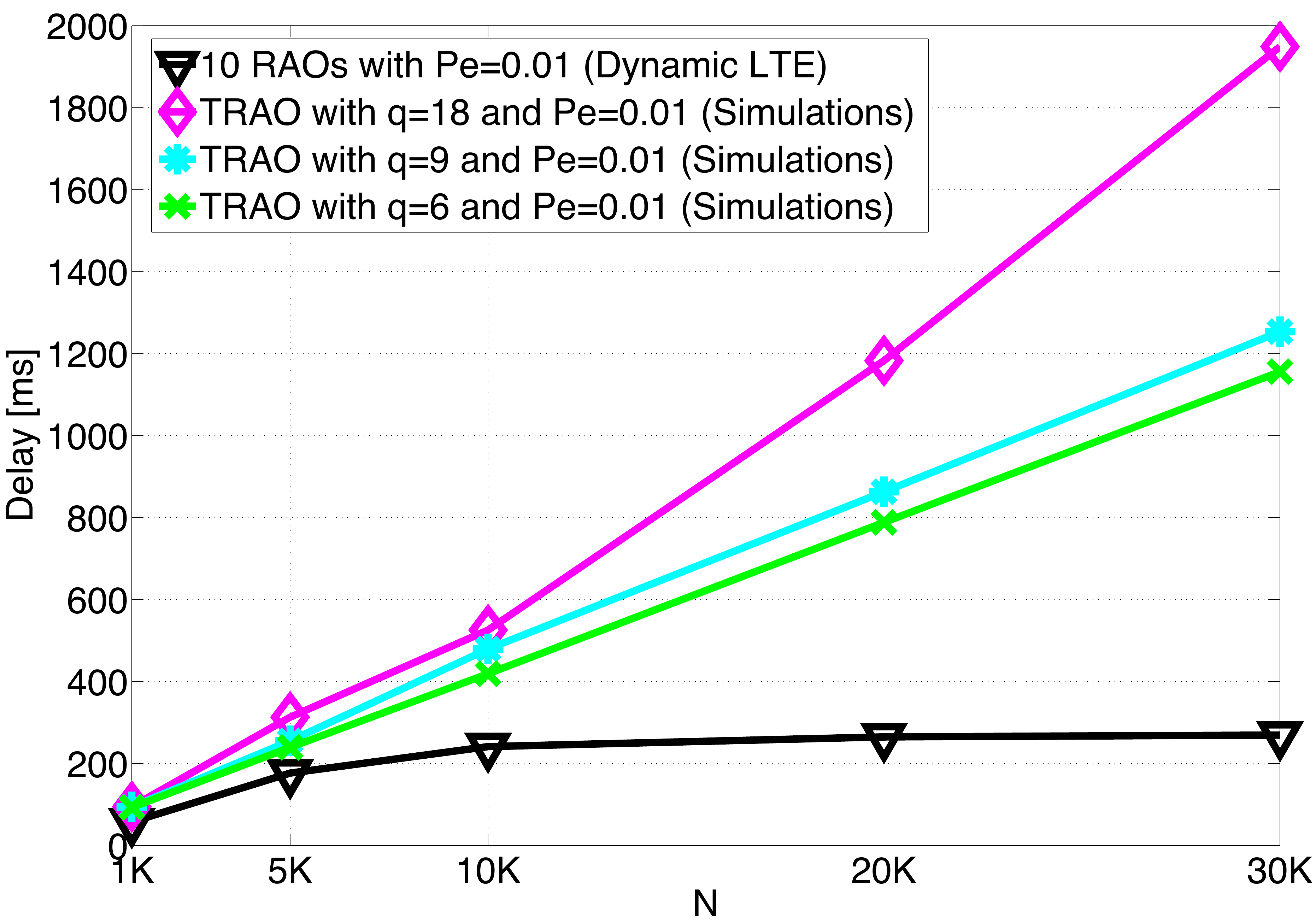}\caption{Average delay experienced by resolved devices.}\label{delay}
	\end{figure}

	The average access delay of devices not in outage is shown in Fig.~\ref{delay}.
	Obviously, this delay is larger for higher $q$, even though the number of required transmissions is lower, see Fig.~\ref{transmissions}.
	 This is due to the fact that for higher $q$ less contention frames $G$ fit in a TRAO, and therefore more TRAOs are needed on average to provide contention frames for the collision resolution.
	We emphasize that the average delay shown for the dynamic allocation applies only to a small percentage of the devices that are not in outage, c.f. Fig.~\ref{fig:outage}.
	
\begin{figure}[t]
	  \centering
	    \includegraphics[width=0.82\columnwidth]{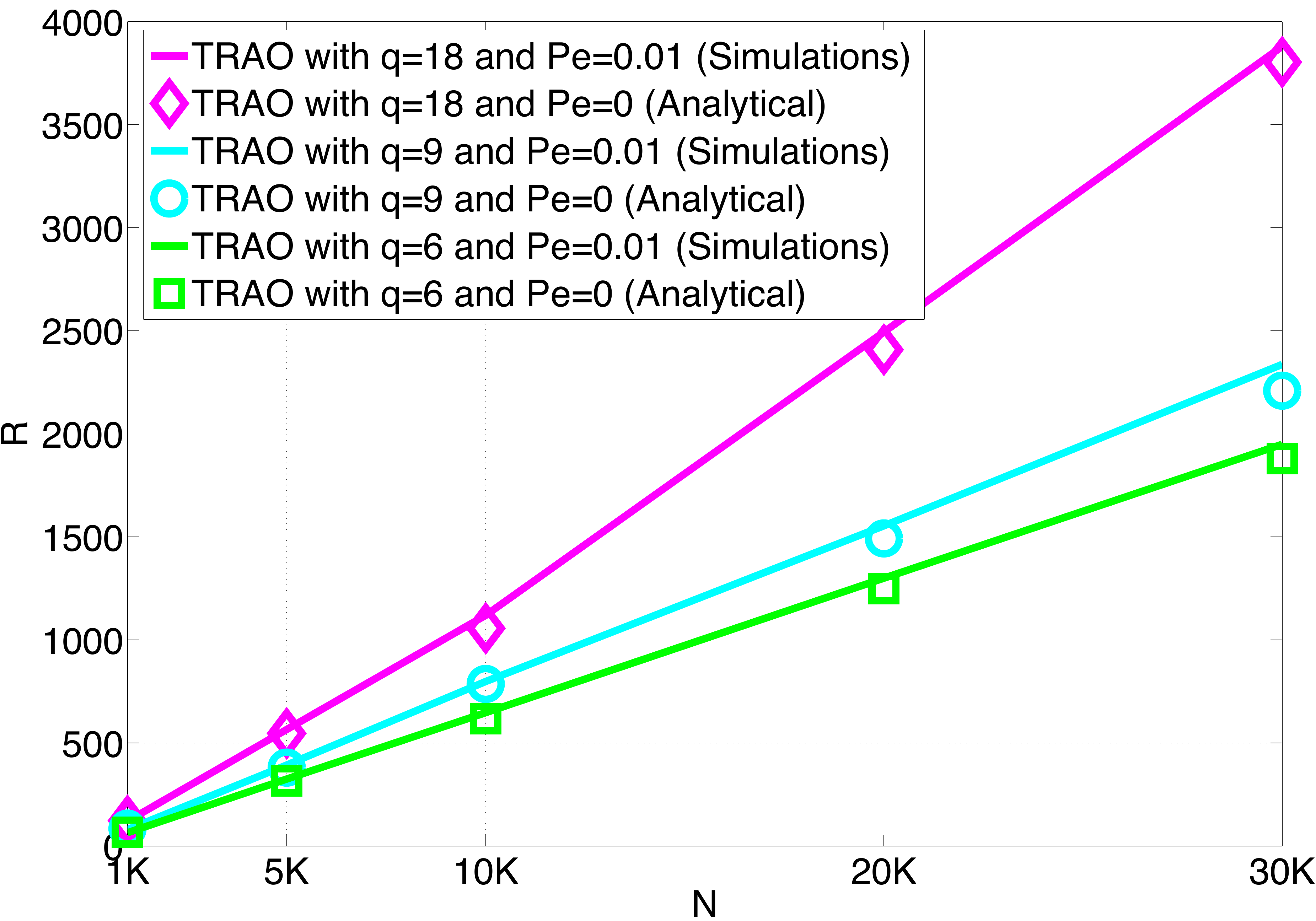}\caption{Average number of TRAOs required.}\label{TRAOs}
	\end{figure}

	The average number of TRAOs  required to resolve \emph{all} the devices $R$ using the proposed scheme is depicted in Fig.~\ref{TRAOs}.
	Obviously, increasing $q$ increases $R$; this can be also inferred from the combination of  \eqref{eq:C} and \eqref{eq:R}.	
		
\begin{figure}[t]
	  \centering
	    \includegraphics[width=0.80\columnwidth]{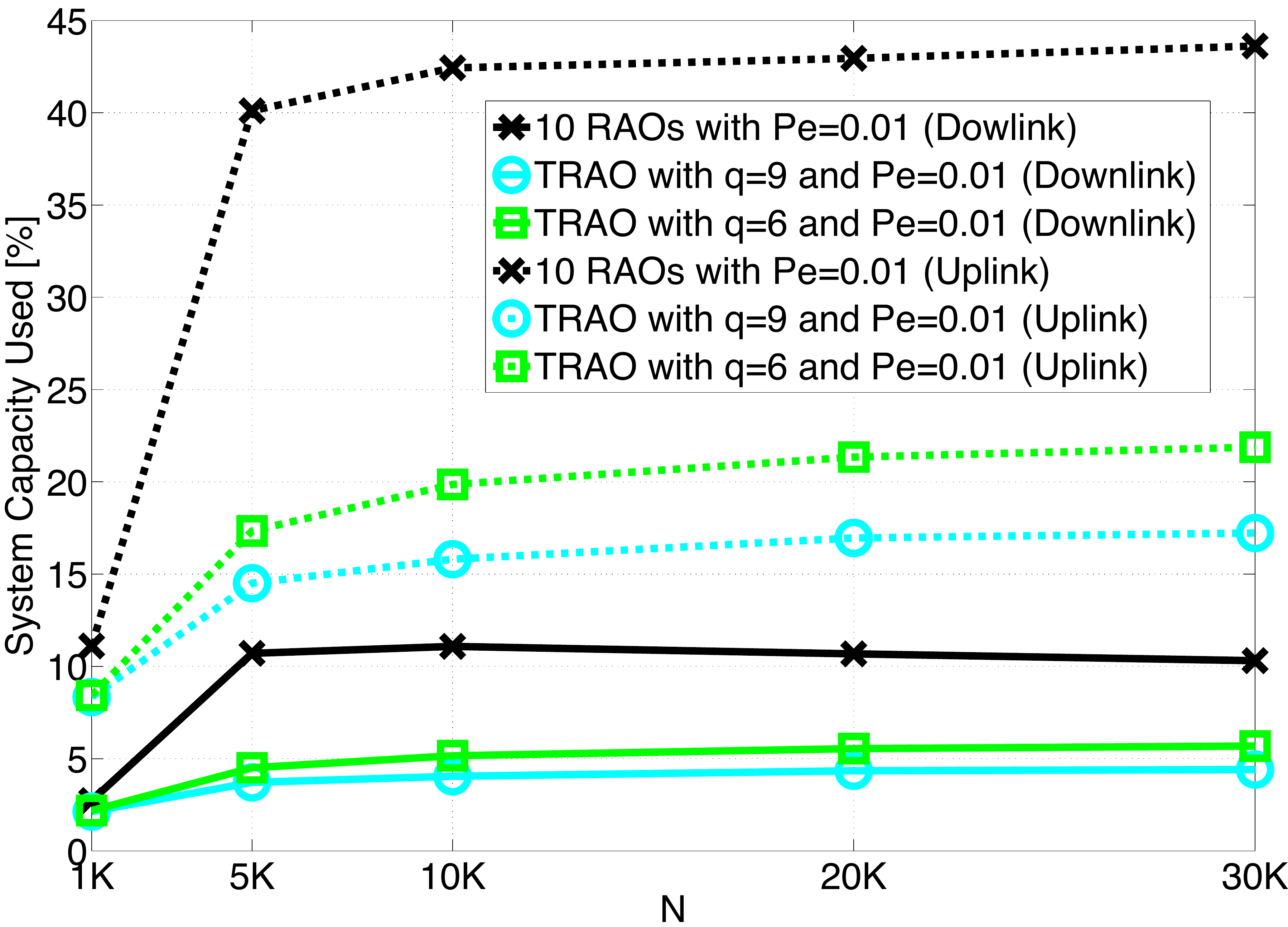}
	    \caption{Percentage of the system capacity for the downlink and uplink used to resolve the devices.}
	    \label{systemCapacity}
	\end{figure}

	Finally, the fraction of the resources used for uplink and downlink for the random access procedure is depicted in Fig.~\ref{systemCapacity}.
	For the downlink, we consider the amount of RBs used to transmit all the required MSG~2, MSG~4 and MSG~4b.
	For the uplink, we consider the amount of RAOs and TRAOs (6~RBs) together with MSG3 (1~RB).
	Obviously, the proposed scheme is significantly less demanding than the dynamic allocation, requiring roughly half of the resources both in the downlink and in the uplink.
	Moreover, we note that these resources are also much more efficiently used, as only an insignificant portion of devices ends in outage, see Fig.~\ref{fig:outage}. 

\section{Conclusion}\label{sec:conclusion}


	In this paper we demonstrated that the LTE RACH becomes easily overloaded with excessive collisions in case of massive synchronous arrivals.
	We also proposed a scheme to deal with such arrivals, which actively pursues collision resolution instead of trying to avoid them.
	The scheme is tailored for LTE RACH and requires only modest modifications of the standard protocol, above the physical layer.
	We demonstrated that the proposed scheme provides reliable and timely service for high numbers of synchronously accessing devices, while requiring less amount of resources than competing schemes.
	Particularly, an astounding 30k devices can be resolved with negligible outage with an average of 5 preamble transmissions and delay of $1.2$ seconds, under realistic probability of transmissions error both in the downlink and in the uplink.
	
	Finally, we note that the proposed scheme allows for efficient and fast delivery of the devices' connection requests, enabling their processing and inspection by the eNodeB.
	In turn, this could provide an extensive basis for the eNodeB to gain insight in the event(s) that caused the massive synchronous arrivals, filter the redundant connection requests during the critical period, and thus alleviate the requirements for the subsequent data stage.

\section*{Acknowledgement}

The research presented in this paper was supported by the Danish Council for Independent Research (Det Frie Forskningsr{\aa}d), grant no. 11-105159 ``Dependable Wireless Bits for Machine-to-Machine (M2M) Communications'' and grant no. DFF-4005-00281 ``Evolving wireless cellular systems for smart grid communications''.

\bibliographystyle{IEEEtran}
\bibliography{bib}

\end{document}